\documentclass[aps,preprint,prl,superscriptaddress,showpacs,amsmath,amssymb]{revtex4}

\usepackage{graphicx,color}
\usepackage{dcolumn}
\newcolumntype{.}{D{.}{.}{-1}}
\usepackage{bm}
\usepackage{epstopdf}
\usepackage{subfigure}

\newcommand{\ala}{$\alpha_{\rm a}$}

\newcommand{\alc}{$\alpha_{\rm c}$}

\newcommand{\tc}{$T_{\rm C}$}
\newcommand{\tsr}{$T_{\rm SR}$}
\newcommand{\tsro}{$T_{\rm SR}^{\rm onset}$}
\newcommand{\cp}{$c_{\rm p}$}

\newcommand{\euz}{Eu$^{2+}$}

\newcommand{\er}{Er\(_2\)CuSi\(_3\)}
\newcommand{\ce}{Ce\(_2\)CuSi\(_3\)}
\newcommand{\nd}{Nd\(_2\)CuSi\(_3\)}
\newcommand{\pr}{Pr\(_2\)CuSi\(_3\)}
\newcommand{\re}{Ln\(_2\)CuSi\(_3\)}
\newcommand{\eu}{Eu\(_2\)CuSi\(_3\)}
\newcommand{\la}{La\(_2\)CuSi\(_3\)}
\newcommand{\cpm}{$c_{\rm p}^{\rm magn}$}

\begin{document}

\title{Magnetic phase diagram of Eu$_2$CuSi$_3$ derived from thermal expansion and magnetostriction studies}

\author{L.~Wang}\email[Email:]{l.wang@kip.uni-heidelberg.de}
\affiliation{Kirchhoff Institute of Physics, Heidelberg University, INF 227, 69120 Heidelberg, Germany}
\author{L.~Wallbaum}
\affiliation{Kirchhoff Institute of Physics, Heidelberg University, INF 227, 69120 Heidelberg, Germany}
\author{C.~Koo}
\affiliation{Kirchhoff Institute of Physics, Heidelberg University, INF 227, 69120 Heidelberg, Germany}
\author{C. D.~Cao}
\affiliation{Department of Applied Physics, Northwestern Polytechnical University, Xi’an 710072, P.R. China}
\author{W.~L\"{o}ser}
\affiliation{Leibniz Institute for Solid State and Materials Research IFW Dresden, Helmholtzstr. 20, 01069 Dresden, Germany}
\author{R.~Klingeler}
\affiliation{Kirchhoff Institute of Physics, Heidelberg University, INF 227, 69120 Heidelberg, Germany}
\affiliation{Center for Advanced Materials, Heidelberg University, INF 225, 69120 Heidelberg, Germany}


\date{\today}

\begin{abstract}
Precise thermal expansion and magnetostriction studies of the intermetallic compound \eu\ in the temperature range between 5 and 300 K are presented. A clear sign of magnetic second order phase transition at the ferromagnetic Curie temperature \tc\ = 34~K indicates strong magneto-structural coupling. Uniaxial thermal expansion data show a large anisotropy between the magnetic easy $c$-axis and the hexagonal $ab$-plane, which is associated with a strongly anisotropic magnetic coupling. A spin-reorientation regime as well as a non-uniform energy scale are indicated by Gr\"{u}neisen analysis of the data at low temperatures. Anomalous contributions to the thermal expansion imply ferromagnetic correlations far above in zero magnetic field. The magnetic phase diagram is constructed, showing the evolution of short- and long-range magnetic order in magnetic fields up to 15~T. The anisotropic magnetic field effect yields anomalous magnetostrictive effects up to about 200~K.

\end{abstract}

\maketitle

\section{Introduction}

Ternary intermetallic compounds \re\ (Ln = rare earth) are of considerable interest because of their diverse types of magnetic ordering and associated complex interplay of structural, orbital, electronic, and magnetic degrees of freedom.~\cite{Hwang96,Tien97,Tien98,Nakamato,Shiokawa,Li,Majumdar99,Cao2008,Cao2010,Zhang2017} Most of the known members of this class crystallize in a hexagonal AlB$_2$-type structure (space group P6/mmm) with random distribution of Cu and Si atoms on the B positions.~\cite{Raman} One exception is \er , crystallizing in a tetragonal ThSi$_2$-type structure.~\cite{Raman} The great diversity of magnetic ground states originates from correlations of the localized $f$-electrons of the rare-earth elements, which are coupled by the conduction electrons according to the Ruderman-Kittel-Kasuya-Yosida theory. Some compounds exhibit unusual transport properties, namely large negative magnetoresistance has been found in \eu\ above the Curie temperature\cite{Majumdar99,Cao2010}, even up to $T \sim 100$~K, and a typical Kondo behavior has been found in \ce\ \cite{Hwang96,Nakamato}. Unusual mass enhancement of \ce\ \cite{Hwang96} and \pr\ \cite{Tien97} has been unveiled by studies on the electronic specific heat. Weak spin glass behavior coexisting with long range ferromagnetic order is observed in \pr\ and \nd .~\cite{Tien97,Li} However, due to particular challenges in single crystal growth \cite{Cao2011}, anisotropic properties have been reported so far only for \ce\ \cite{Nakamato} and \eu\ \cite{Cao2010} while all other experimental data rest on polycrystalline samples.

Due to its half-filled shell in the 4$f^7$ electronic configuration, \eu\ has a particular role among the lanthanides-based series of ternary intermetallics. The Eu$^{2+}$-ions in \eu\ do not follow the lanthanide contraction rule.~\cite{Majumdar99,Cao2010} Instead, unit cell volume and lattice parameters in \eu\ are abnormally enlarged, similar to what is observed in the more frequently studied family R$_2$PdSi$_3$.~\cite{Mallik} Ferromagnetic order in \eu\ evolves at \tc\ = 34~K. Above \tc , magnetization data show a magnetic moment of $(7.8\pm 0.1)\mu_B/\text{Eu}$, i.e., a stable valence of 2+ of Eu-ions. The anisotropy of the electron spin resonance signal of divalent \euz -ions proves appreciable short-range ferromagnetic correlations up to $T\sim 100$~K, i.e., far above \tc .~\cite{Cao2010} Below \tc , there is a sizeable magnetic anisotropy with the crystallographic $c$-axis being the easy magnetic axis. In the temperature regime around 10 to 20~K the magnetic anisotropy gradually vanishes most probably due to a spin reorientation. It may be expected that, similar to other intermetallic materials (e.g., HoNi$_2$B$_2$C~\cite{Schneider}, Tb$_x$Gd$_{1-x}$Al$_2$~\cite{Moral1986}, Mo$_{5+y}$Si$_{3−y}$B$_x$~\cite{Zhao2004}), that magnetic anomalies are tightly coupled with lattice changes. However, no investigations of magneto-structural effects are known for \eu\ or any other member of this class of materials. Therefore, the investigation of thermal expansion and magnetostriction on a \eu\ single crystal will elucidate the interrelation of structural, magnetic, and electron degrees of freedom. In the present work, we investigate in detail of the uniaxial lattice distortions induced by the magnetic transition as well as the influence of external magnetic fields. A detailed magnetic phase diagram of \eu\ focusing on the low temperature range up to 70 K is derived from the data.

\section{Experimental}

\eu\ single crystals have been grown by the traveling-solvent floating-zone method as described in Ref.~\onlinecite{Cao2011}. The relative length changes $\Delta L_i/L_i$ along the crystallographic $c$- and $a$-directions were studied (as the $a$- and $b$-axes are crystallographically equivalent) on a cuboidal shaped crystal which dimensions are 1.831~mm ($\|a$), 3.408~mm ($\|b$), and 1.792~mm ($\|c$). The measurements were done by means of a three-terminal capacitance dilatometer.~\cite{Wang2009} For investigating the effect of magnetic fields, the length changes and the thermal expansion coefficients $\alpha_i = 1/L_i\cdot dL_i(T)/dT$ were studied in magnetic fields up to 15~T applied along the axis $i$, respectively. In addition, the field induced length changes $\Delta L_i(B)/L_i$ were measured at temperatures $T$ from 5~K to 200~K in magnetic fields up to 15~T and the magnetostriction coefficients $\lambda_i = 1/L_i\cdot dL_i(B)/dB$ were derived.

\section{Thermal expansion and Gr\"{u}neisen scaling}

\begin{figure}[tb]
\includegraphics [width=0.95\columnwidth,clip] {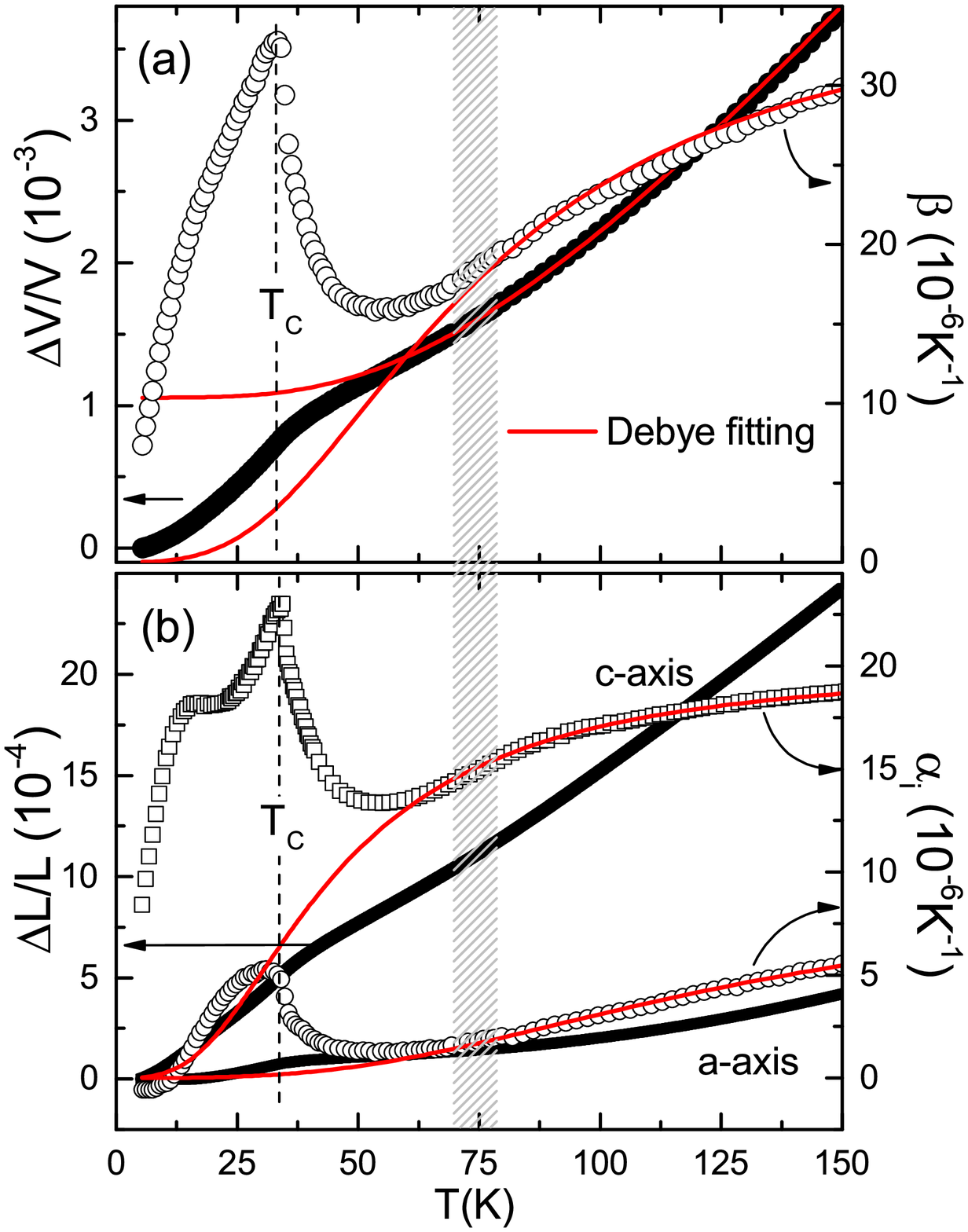}
\caption{(a) Relative volume changes $\Delta$L/L (left ordinate) and volume thermal expansion coefficient $\beta$ of \eu\ vs. temperature $T$. \tc\ denotes the Curie temperature at which the Eu-moments order ferromagnetically. The red lines are a Debye fit to the data well above \tc\ (see the text).  (b) Relative length changes $\Delta$L/L (left ordinate) and linear thermal expansion coefficients $\alpha$ of \eu\ measured within the $a$/$b$-plane (circles) and along the $c$-axis (squares) vs. temperature $T$. } \label{Fig1}
\end{figure}

The temperature dependence of the uniaxial length changes in \eu\ shows that magnetic order is associated with pronounced magnetoelastic effects (Fig.~\ref{Fig1}b) both along $c$ and in the $ab$-plane. Accordingly, the volume significantly shrinks upon evolution of magnetic order and the volume thermal expansion coefficient $\beta = 2\alpha_a + \alpha_c$ displays a peak (Fig.~\ref{Fig1}a). Qualitatively, this implies positive hydrostatic pressure dependence d\tc /d$p$. In addition, there is a broad shoulder in $\beta$ at around 15~K and the region of anomalous volume changes extends well above \tc . According to Ref.~\onlinecite{Lindbaum94}, the electronic ($\beta_{el}$) and phononic ($\beta_{ph}$) contributions to the thermal expansion coefficient can be described by

\begin{align}
\beta_{el}&=K_1\cdot T^2 \\
\beta_{ph}&=K_2T\cdot D(\Theta_D/T)
\end{align}

with $K_1$ and $K_2$ being constants and $D$ being the Debye function. The red solid lines in Fig.~\ref{Fig1} show the results of the fitting procedure to the experimental data well above \tc. For the fitting of both the volume and the uniaxial expansion data, the same Debye temperature $\Theta_D = 425$~K independently obtained from analysing the specific heat of \la\ (data from Ref.~\onlinecite{Hwang96}) is used here to ensure the consistency of the procedure.~\footnote{Note, that our analysis of \cp\ of \la\ applying the Debye function up to 50~K yield a different value for $\Theta_D$ as compared to the $T^3$-law fitting used in Ref.~\onlinecite{Hwang96}. The obtained $\Theta_D=445$~K was renormalised in order to account for the different masses.} Quantitatively, the differences between the estimated non-magnetic lengths changes and the actual experimental data at $T=5$~K amounts to $\Delta V/V \approx 1\cdot 10^{-3}$. The temperature range of this divergence is clearly visible in $\beta$ which signals anomalous contributions to the volume changes up to $\sim$75~K. As will be shown below, the anomalous length changes $\Delta L'_i=\Delta L_i-\Delta L_i^{\rm el,ph}$ are strongly suppressed by external magnetic fields which suggests its magnetic nature. Short-range ferromagnetic correlations have been indeed detected in \eu\ by electron spin resonance below $T \simeq 100$~K.~\cite{Cao2010} We conclude that the observed anomalous length changes are associated to ferromagnetic short range magnetic correlations.

\begin{figure}[tb]
\includegraphics [width=0.95\columnwidth,clip] {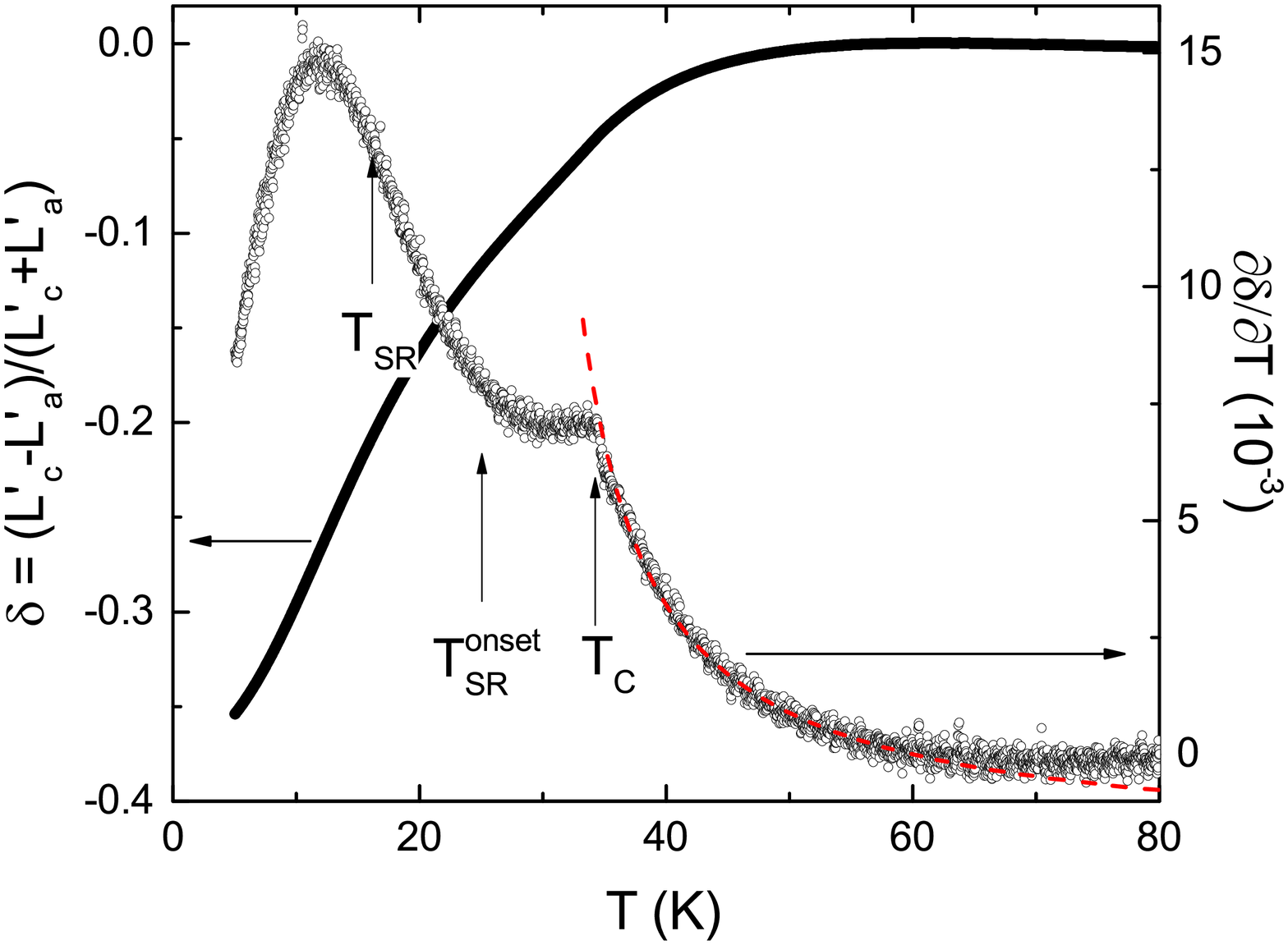}
\caption{Distortion parameter $\delta$ and its derivative as derived from the anomalous lengths changes $\Delta L'_i$. \tsr\ indicates the spin reorientation temperature. The dashed line is a phenomenological Curie-Weiss fit to the data.} \label{OP}
\end{figure}

The uniaxial length changes in \eu\ along the $a$- and the $c$-axis, respectively, show pronounced anisotropy which is reflected by the fact that, well above the anomalies at $T = 150$~K, \alc\ is about three times larger than \ala\ (Fig.~\ref{Fig1}b). As in $\beta$, in both directions the onset of long range magnetic order is associated with clear anomalies. Interestingly, in both directions long range magnetic order is associated with shrinking of the respective lattice parameters $a$ and $c$, i.e., positive uniaxial pressure dependence of \tc . In \alc , there is a peak at \tc\ followed by a broad hump centered around 15~K. In contrast, the anomalous contributions are much broader in \ala\ and there is only a rather jump-like anomaly at \tc . Upon further cooling, no additional hump can be identified. The structural changes are further illustrated by the distortion parameter $\delta = (L'_c-L'_a)/(L'_c+L'_a)$ and its derivative shown in Fig.~\ref{OP}. The data confirm that the ferromagnetic phase transition at \tc\ is associated with a kink in the distortion parameter and a large regime of structural fluctuations up to about 75~K. Upon cooling towards \tc , $\partial\delta/\partial T$ smoothly increases in a Curie-Weiss-like manner, i.e., as $C/(T+\Theta)$, with $\Theta\approx 27.5$~K. Note, that this description of the derivative of the parameter $\delta$ does not reflect the high-temperature mean-field approximation but only phenomenologically describes the behavior upon approaching \tc . Just below \tc , $\partial\delta/\partial T$ is temperature independent, i.e., $\delta$ decreases linearly. A large hump of $\partial\delta/\partial T$ centered at 12~K is observed in the temperature regime below about 25~K. These distortions appear in the temperature regime where magnetic anisotropy starts to decrease through a spin reorientation process which finally yields vanishing anisotropy below $T\sim 5$~K.~\cite{Cao2010} To be specific, upon cooling magnetic moments rotate from the easy magnetic $c$-axis towards $\perp c$, thereby lifting the magnetic anisotropy. One may attribute the low temperature humps of the specific heat\cite{Cao2010} and of $\partial\delta/\partial T$ as the thermodynamic signatures of this spin-reorientation towards the $ab$-plane.


\begin{figure}[htb]
\includegraphics [width=1.0\columnwidth,clip] {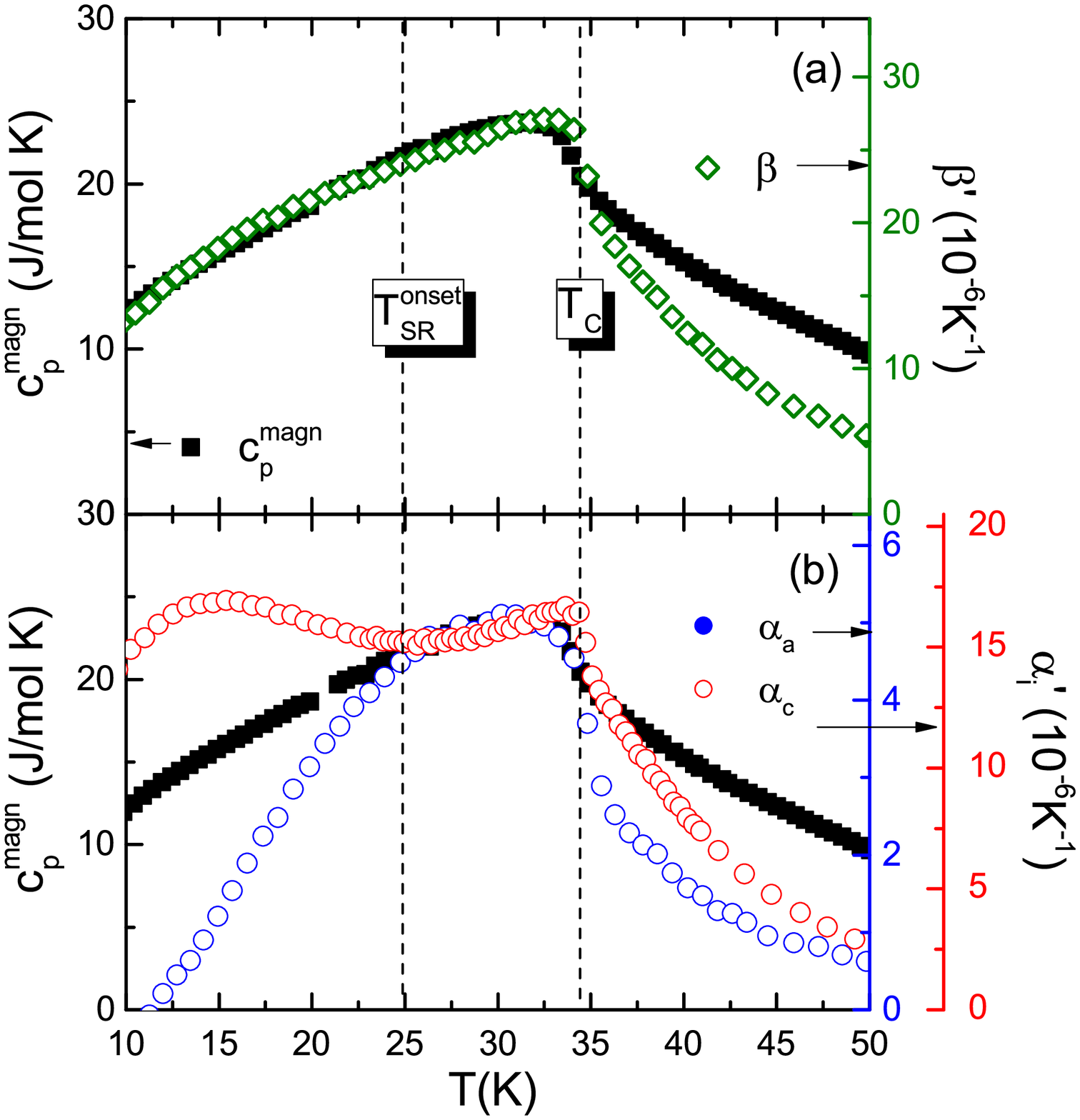}
\caption{(a) Anomalous volume thermal expansion coefficient $\beta$' and magnetic specific heat \cpm\ illustrating Gr\"{u}neisen scaling in \eu . (b) Anomalous linear thermal expansion coefficients \ala ' and \alc ' , and \cpm . } \label{Fig3}
\end{figure}

In order to discuss the anomalous length changes in more detail, Fig.~\ref{Fig3} presents the low temperature thermal expansion and the specific heat~\cite{Hwang96}. Again, the data have been corrected for the phononic and electronic contributions obtained from analysing of \cp\ of \la , yielding the magnetic contributions $\beta$' and $\alpha_i$'.~\cite{Hwang96,Cao2010} Different ordinate scales have been used to highlight the similarities and differences in the temperature dependencies. The data show that, at temperatures $\lesssim$~\tc , the magnetic specific heat \cpm\ resembles quite well the volume thermal expansion coefficient $\beta$' while scaling fails at $T>$~\tc\ (Fig.~\ref{Fig3}a). At temperatures $T>$ \tc , our data exclude similar temperature dependencies of $\beta$' and \cpm .

In the presence of one dominant energy scale $\epsilon$ of the respective ordering phenomenon, the scaling between \cpm\ and $\beta$' is described by the Gr\"{u}neisen relation

\begin{equation}
\frac{\beta'}{c_p^{\rm magn}}=\frac{1}{V}\left. \frac{\partial \ln \epsilon}{\partial p}\right|_T.\label{eqgruen}
\end{equation}

Applying Eq.~\ref{eqgruen} to the data in Fig.~\ref{Fig3}a yields the hydrostatic pressure dependence $\partial \ln \epsilon /\partial p = 13(2)\cdot 10^{-2}$/GPa for the temperature range $T \leq$~\tc\ where \cpm\ scales to $\beta$'. Assuming $\epsilon$ being proportional to \tc , this yields $\mathrm{d}T_\mathrm{C}/\mathrm{d}p_a = 4.4(9)$~K/GPa.

In contrast to the volume effect, there is no general Gr\"{u}neisen scaling below \tc\ if the uniaxial pressure effects are considered. However, \cpm\ still fairly scales to both \ala ' and \alc ' in the reduced temperature regime 25~K $\leq T \leq$~\tc\ (see Fig.~\ref{Fig3}b). Here, the analysis yields $\partial \ln \epsilon /\partial p_a = 2.5(2)\cdot 10^{-2}$/GPa and $\partial \ln \epsilon /\partial p_c = 8.0(9)\cdot 10^{-2}$/GPa  or $\mathrm{d}T_\mathrm{C}/\mathrm{d}p_a = 0.85(10)$~K/GPa and $\mathrm{d}T_\mathrm{C}/\mathrm{d}p_c = 2.7(3)$~K/GPa, respectively. For both \alc ' and \ala ', simple Gr\"{u}neisen scaling can be excluded for $T \leq$ 25~K.

In general, one may identify three temperature regions with different behavior as indicated in Fig.~\ref{Fig3}: (1) At $T>$ \tc , the anomalous volume changes as well as the uniaxial length changes do not obey Gr\"{u}neisen scaling. (2) Upon evolution of long range magnetic order at \tc , there is a rather temperature independent hydrostatic pressure dependence of $\epsilon$ down to 5~K. For the uniaxial pressure dependencies, this regime is restricted to \tsro\ $\leq T\leq$~\tc . (3) Below \tsro , where the distortion parameter $\delta$ signals spin reorientation, the uniaxial pressure dependencies for pressure applied along the $a$- and $c$-axis, respectively, strongly differ.


\section{Effect of external magnetic fields}

\begin{figure}[tb]
\includegraphics [width=1.05\columnwidth,clip] {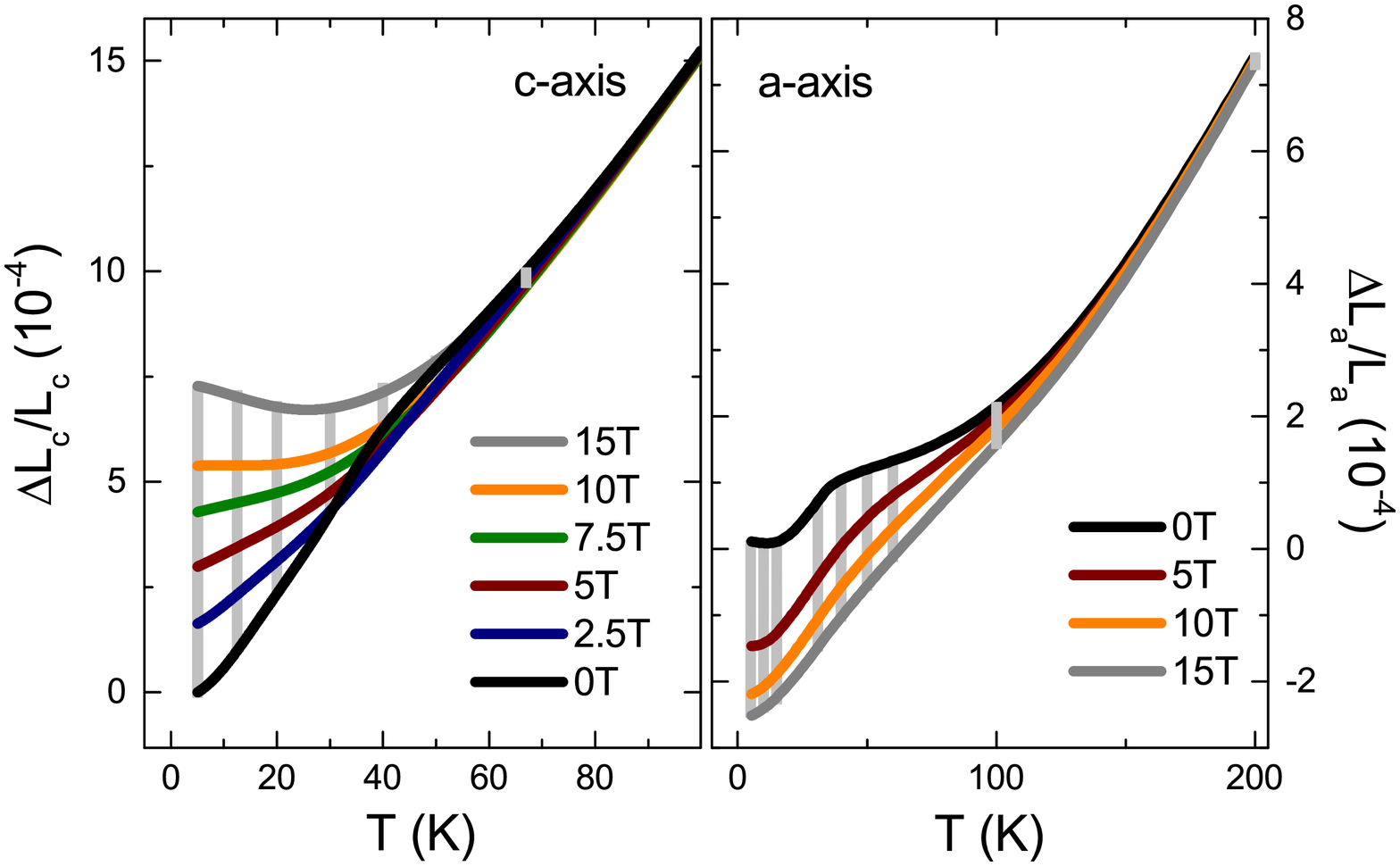}
\caption{Relative length changes $\Delta$L/L in different external magnetic fields vs. temperature measured along the $c$- and $a$-axis, respectively. The data at $B\neq 0$ have been shifted according to the measured magnetostriction at 5~K. The grey vertical lines show the measured magnetostriction up to 15~T at various temperatures (see Fig.~\ref{LvsB}).} \label{LvsT}
\end{figure}

\begin{figure}[tb]
\includegraphics [width=0.95\columnwidth,clip] {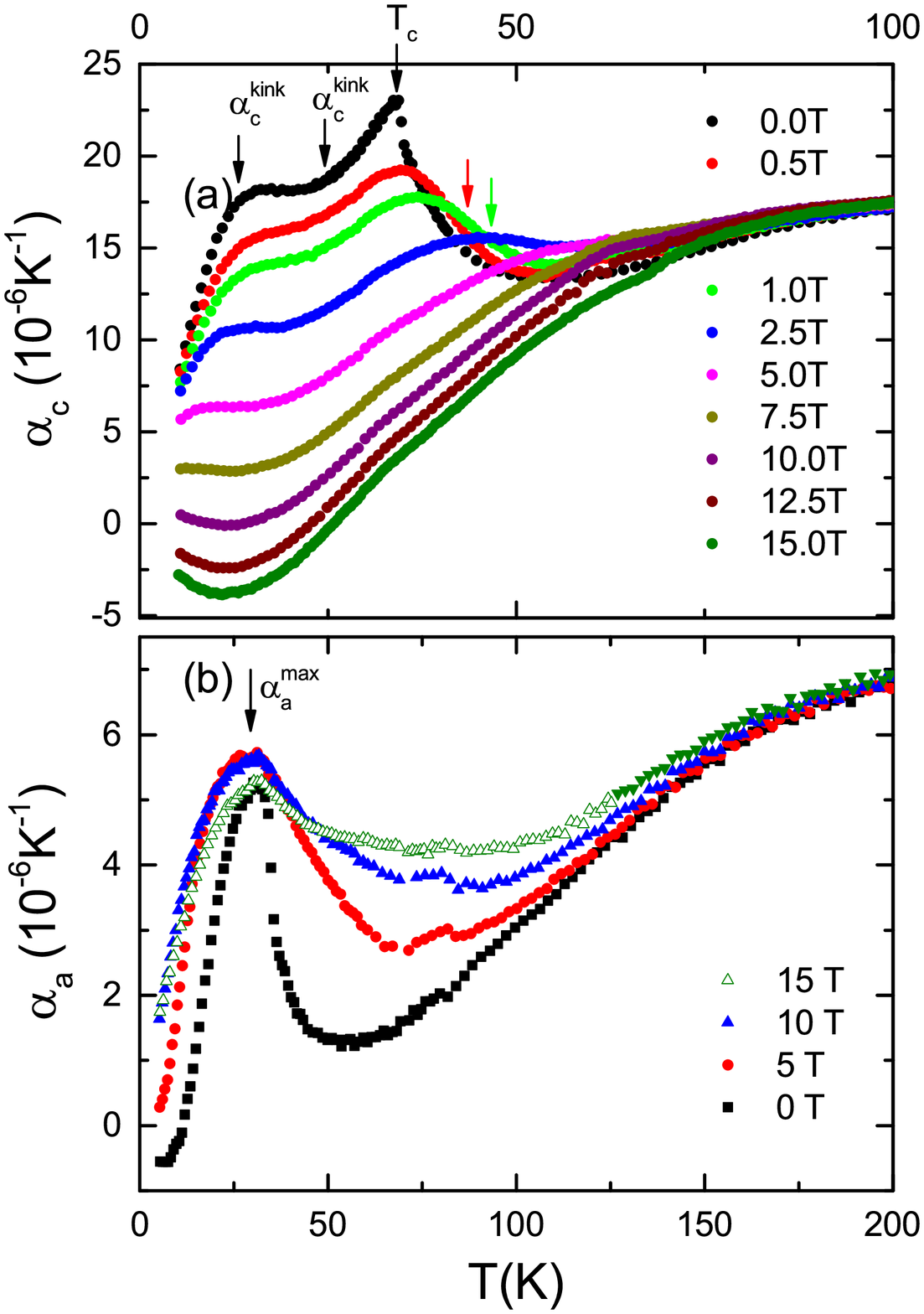}
\caption{Linear thermal expansion coefficients $\alpha_c$ (a) and $\alpha_{a}$ (b) of \eu\ in different external magnetic fields applied in the direction of measurements. Arrows indicate anomalies as discussed in the text.} \label{alphafeld}
\end{figure}

Figures~\ref{LvsT} and~\ref{alphafeld} show the effect of external magnetic fields on the uniaxial length changes. Magnetic fields up to 15~T affect the lengths changes well above \tc\ indicating structural changes up to $\sim 80$~K for $B\|c$ and $\sim 200$~K for $B\|a$.
While the $a$-axis shrinks in external magnetic field in the entire temperature range up to 200~K, the $c$-axis displays a heterogeneous response and anomalous magnetostriction is restricted to $T<100$~K. Upon application of $B=15$~T, the $c$-axis overall shrinks above $T\simeq 60$~K while it strongly increases below. As will be discussed in more detail below, there is a sign change of the magnetostriction coefficient $\lambda_c(B)$ at \tc\ $< T \lesssim 75$~K.

The field effect on the anomaly at \tc\ = 34~K is most visible if the thermal expansion coefficients in Fig.~\ref{alphafeld} are considered. Both \ala\ and \alc\ show a broadening of the anomaly at \tc\ in external magnetic field. The ferromagnetic nature of the phase transition is reflected by the strong shift of anomalous length changes to higher temperatures and the sharp anomalies observed at $B=0$ are not observed in finite magnetic field. In magnetic fields $B \geq 5$~T, no clear anomaly can be observed anymore. Even in finite fields $B<5$~T, the $\lambda$-like peak in \ala\ which is the most distinct signature of the phase transition smears out and converts into a jump-like feature. We hence consider the middle of this jump being a relevant signature for the evolution of ferromagnetic correlations. Together with all other features extracted from the data as described in the following, it is used to construct the magnetic phase diagram in Fig.~\ref{phd}. In contrast to the anomaly at \tc , the low-temperature hump indicative of spin-reorientation from the $c$-axis into the $ab$-plane is suppressed in external magnetic fields. This is straightforwardly attributed to the fact that the spin reorientation towards $\perp c$ is hindered by $B\|c$. We define this spin-reorientation regime using the concave kink in the curve, and the convex kink is used to define the crossover to isotropic magnetic behavior, marked by the black arrows in Fig.~\ref{alphafeld}a. For fields larger than 10~T, \alc\ becomes negative at low temperature and the features signaling spin-reorientation are not observed in the temperature regime under study. In the $ab$-plane, the effects of magnetic fields on $\alpha$ are quite different. The anomaly at \tc\ is smeared out as well, but a clear maximum remains observable even in fields up to $B=15$~T where a kink is seen in \ala\ at $T\approx 32$~K.

\begin{figure}[tb]
\includegraphics [width=1.05\columnwidth,clip] {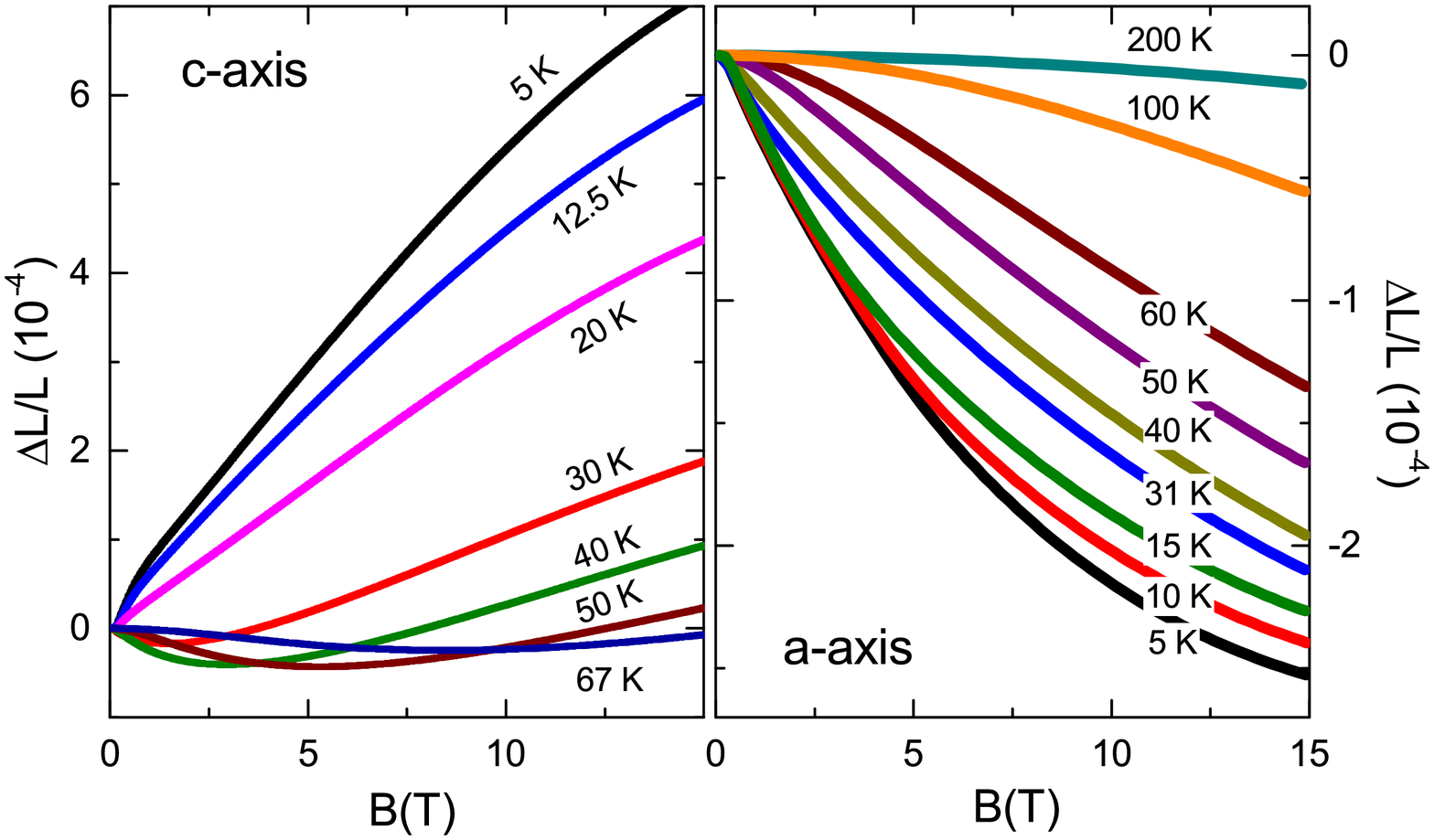}
\caption{Magnetostriction $\Delta$L/L at different temperatures measured along the $c$- and $a$-axis, respectively.} \label{LvsB}
\end{figure}

\begin{figure}[tb]
\includegraphics [width=0.95\columnwidth,clip] {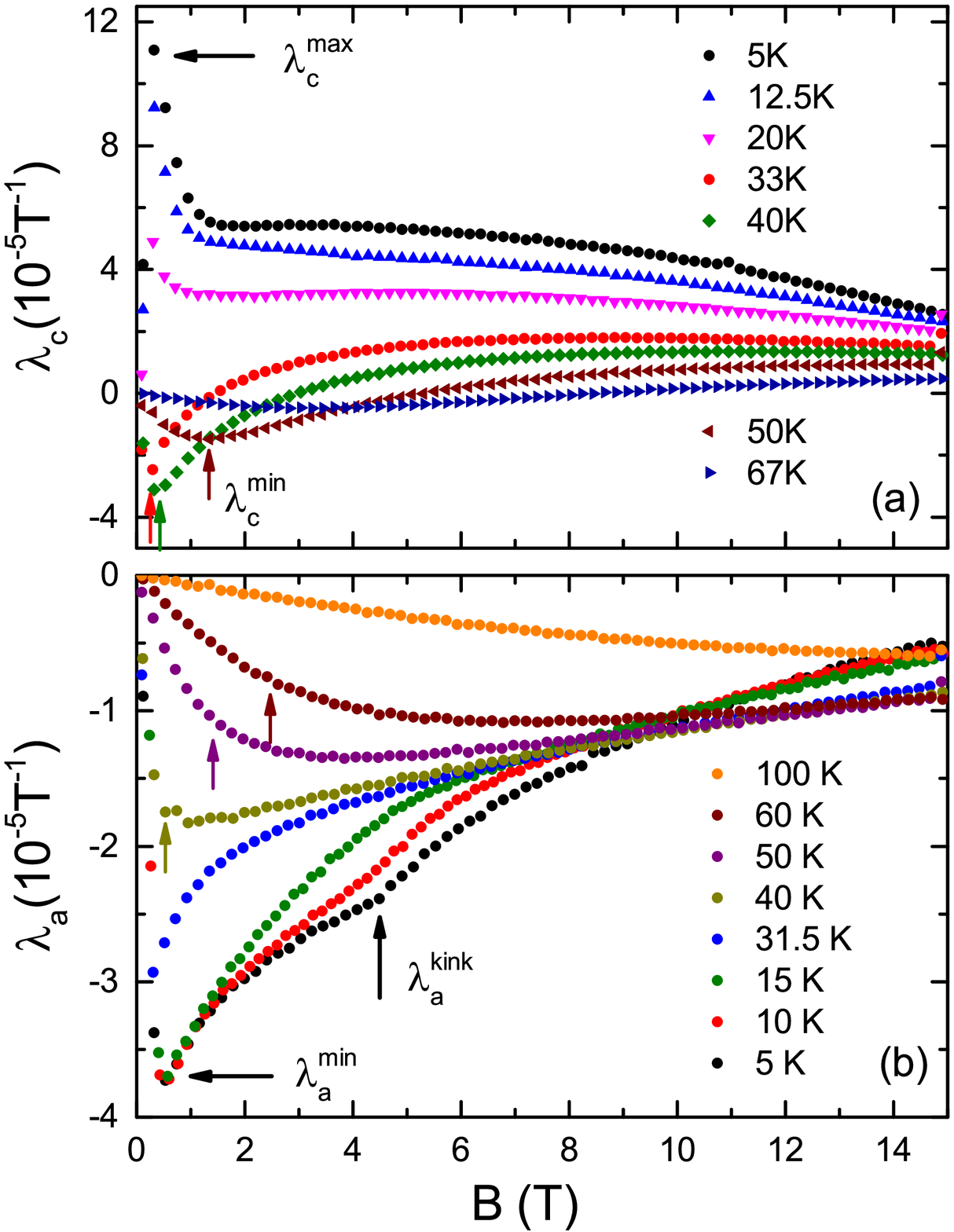}
\caption{(a) Linear magnetostriction coefficients $\lambda_c$ (a) and $\lambda_{a}$ (b) of \eu\ at different temperatures. The arrows highlight anomalies discussed in the text.} \label{ms}
\end{figure}

These differences associated with the actual direction of the magnetic field with respect to the crystallographic axes are also seen in the linear magnetostriction coefficients $\lambda_c$\ and $\lambda_{a}$\ displayed in Fig.~\ref{ms}. In the $ab$-plane, $\lambda$\ is negative at all fields and temperatures under study. At $T=5$~K, there are two anomalies as indicated by arrows in Fig.~\ref{ms}b. One of which is associated with a peak at $B=0.5$~T and the other with a kink at $B=4.5$~T. The position of the peak does not change upon heating up to 15~K, appears at a slightly reduced field at $T=31.5$~K and disappears above \tc. A very similar behavior with however opposite sign is found in $\lambda_a$ which shows a peak maximum at low field. Above \tc , the minimum in $\lambda_a$ converts into a kink where the slope of $\lambda_a$ changes. This kinks becomes more and more broad upon heating. Correspondingly, there is a minimum in $\lambda_c$ above \tc\ which is shifted to higher fields upon heating. At $T=50$~K, there is still a broad anomaly centered around $B=1.5$~T. Both features observed above \tc\ display the same $B$~vs.~$T$ behavior (see Fig.~\ref{phd}) so that we conclude a common nature. In contrast, the above mentioned kink in $\lambda_{a}$\ is restricted $T \leq$ \tsr . The nonuniform response of the length changes for $B\|c$ are displayed by the magnetostriction coefficients obtained at \tc\ $\leq T <$ 100~K. Associated with the minimum discussed above, magnetostriction is negative at low fields and positive at high magnetic fields.

\begin{figure}[tb]
\includegraphics [width=0.95\columnwidth,clip] {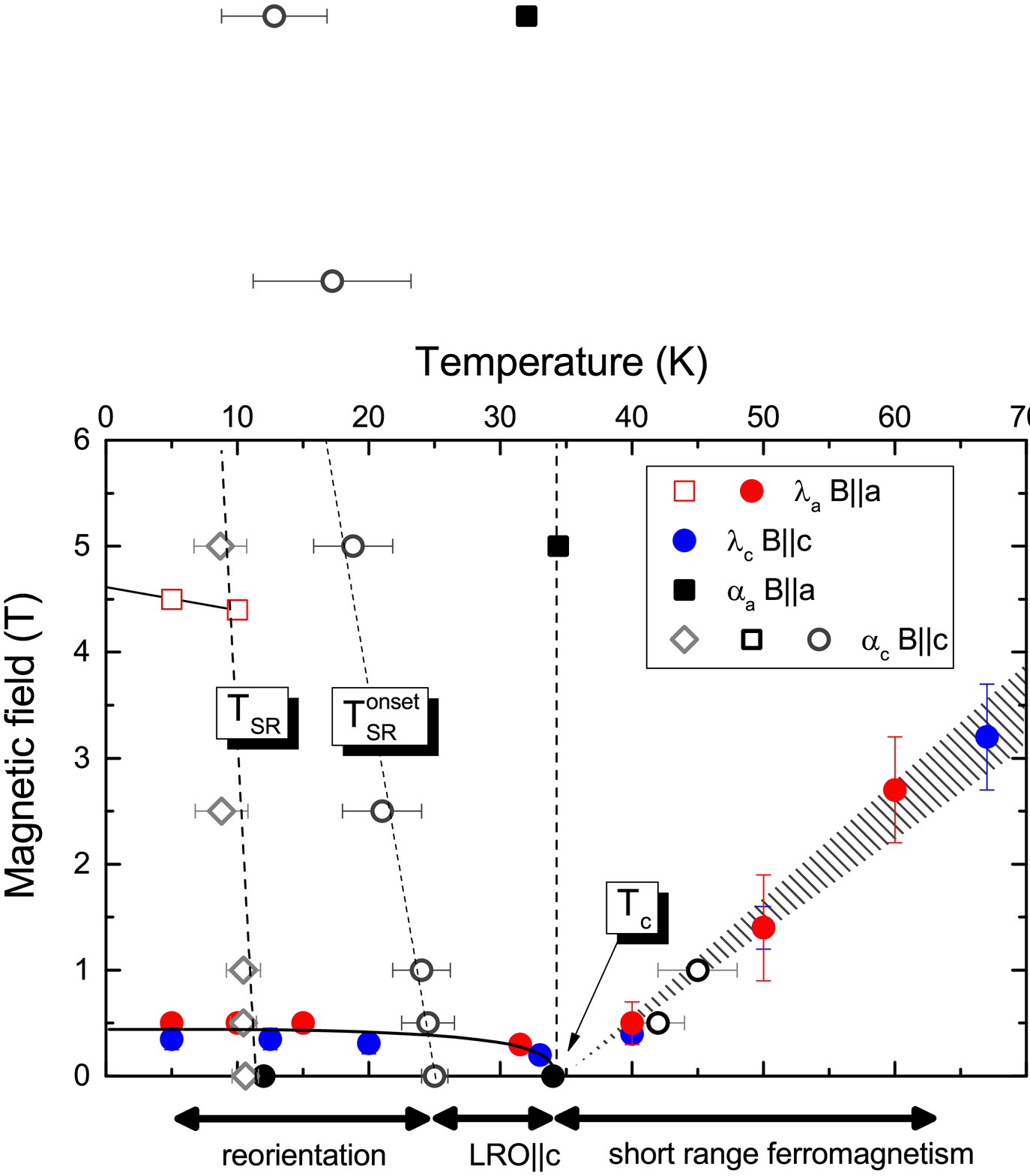}
\caption{Magnetic phase diagram of \eu . Color symbols (red \&\ blue) are deduced from the magnetostriction $\lambda_i$, black symbols (solid \&\ hollow) are from the temperature dependent length changes $\alpha_i$ at different magnetic fields. The black arrows below the graph illustrate regions of different magnetic behaviour at $B=0$. The dashed and shaded lines are guides to the eye indicating the evolution of ferromagnetism and spin reorientation.} \label{phd}
\end{figure}

\section{Discussion and conclusions}

The onset of long range magnetic order yields shrinking of the sample both along the $c$- and the $a$-direction. \eu\ displays stable valence of Eu$^{2+}$ below room temperature. Its 4$f^7$ electronic configuration implies no orbital momentum and the Steven's factors vanish so that magnetic anisotropy and magnetoelastic effects induced by spin-orbit coupling is negligible. In the half-filled shell, crystal field effects can be excluded either. We also note, that the observed anomalous length changes are not associated with unstable Eu valence.~\cite{Neumann1985} The strong magnetoelastic effect might result from either dipole-dipole interaction or anisotropic magnetic coupling. Above \tc , both \alc\ and \ala\ imply the presence of anomalous length changes at elevated temperatures up to $\sim$80~K, i.e., far above the Curie temperature. We attribute these anomalous length changes to the presence of ferromagnetic correlations. In this temperature regime, the evolution of short range ferromagnetic order is associated with a structural distortion, i.e., finite distortion parameter $\delta$. Recent ESR data suggest that the effective internal field in the short-range-correlated region is parallel to the $c$-axis.~\citep{Cao2010} Noteworthy, Gr\"{u}neisen scaling does not apply in this temperature regime. Due to the fact that the Gr\"{u}neisen parameter obtained in the long-range ordered phase does not describe the relation between $\beta$' and \cpm\ at $T>$~\tc , one has to conclude the different nature of short range and long range magnetic order. In particular, the failure of simple Gr\"{u}neisen scaling at \tc\ $< T \leq 50$~K suggests competing ordering phenomena in this temperature regime.

This is corroborated by the observed anisotropic field dependencies at high temperatures. In general, both magnetic fields $\| c$ and $\perp c$ stabilize the ferromagnetic correlations and there is a pronounced magnetic field dependence. Upon application of magnetic fields $B\| c$, the sharp thermodynamic signature of long-range magnetic order at \tc\ disappears and strongly smears out already in small magnetic field. Although this is a typical behaviour for ferromagnets where no true magnetic phase transition appears in finite magnetic field, we note that our data do not unambiguously prove long range order at $B\|c\geq 0.5$~T. The data for $B\perp c$ do not clarify this issue either, as the jump-like feature in \ala\ disappears in magnetic field. However, in contrast to the case $B\| c$, the maximum in \ala\ is not significantly affected as indicated by the vertical dashed line in Fig.~\ref{phd}. The phase boundary evidenced by the magnetostriction data at $B \sim 0.4$~T might hence either indicate melting of (true) long-range order or the stabilisation of a competing ordered phase. A spin-flop nature of this phase boundary can be excluded because it is observed in both field directions as well as above and within the spin-reoriented phase. The origin of the feature at around 4.5~T and $T<$ \tsr\ is unknown. The fact that \tsr\ and the associated onset temperature are suppressed by $B\|c$ support the scenario that, at \tsr , reorientation of the magnetic moments from $\|c$ towards the $ab$-plane appears. For $B\|a$, the data do not allow to extract \tsr .

Despite the easy $c$-axis nature of long range order in \eu , magnetic fields $B\|c$ do stabilize the short range magnetic order less than $B\perp c$. This observation supports our conclusion that the nature of short range order differs from the magnetically ordered state. Our observation of magnetostrictive effects even at 200~K straightforwardly connects to the anomalous magnetoresistance in \eu .~\cite{Majumdar99,Cao2010} Upon application of $B=6$~T, negative magnetoresistive has been observed up to about 80~K for $B\|c$ and about 130~K for $B\perp c$.~\cite{Cao2010} On the one hand, structural changes proven by our high-precision dilatometry data presented here naturally explain associated changes of the electric transport properties. Quantitatively, the larger magnetoresistance for $B\perp c$ observed in the experiment can be explained by the anisotropic effect of magnetic field on the structure, i.e., stronger magnetostrictive effects for $B\perp c$.



%
In summary, we have studied the thermodynamic properties and their field dependence of intermetallic single crystalline \eu . At the ferromagnetic ordering temperature, strong magnetostrictive coupling yields a pronounced $\lambda$-like anomaly in the thermal expansion coefficient. The distortion parameter signals short-range order far above \tc . A spin-reorientation regime as well as a non-uniform energy scale are indicated by Gr\"{u}neisen analysis of the data at low temperatures. The magnetic field effect both on the long-range order, the spin-reorientation, and the short-range order is highly anisotropic and yields anomalous magnetostrictive effects up to about 200~K.

\begin{acknowledgements}

RK gratefully acknowledges fellowship of the Marsilius Kolleg Heidelberg. CDC acknowledges financial support
by the National Natural Science Foundation of China Grant No. 51471135, the National Key Research and Development Program of China under contract No. 2016YFB1100101, and Shaanxi International Cooperation Program.

\end{acknowledgements}

\bibliography{Eu213TE}

\end{document}